\documentclass[%
 aip,
 amsmath,amssymb,
preprint,%
]{revtex4-1}

\usepackage{graphicx}
\usepackage{dcolumn}
\usepackage{bm}

\usepackage[utf8]{inputenc}
\usepackage[T1]{fontenc}
\usepackage{mathptmx}
\usepackage{etoolbox}

\makeatletter
\def\@email#1#2{%
 \endgroup
 \patchcmd{\titleblock@produce}
  {\frontmatter@RRAPformat}
  {\frontmatter@RRAPformat{\produce@RRAP{*#1\href{mailto:#2}{#2}}}\frontmatter@RRAPformat}
  {}{}
}%
\makeatother
\begin{document}

\preprint{}

\title[Individual species chemical potentials in MSA]{Individual ion species chemical potentials in the Mean Spherical Approximation}
\author{Johan S. H{\o}ye}
 \email{johan.hoye@ntnu.no}
 \affiliation{Department of Physics, Norwegian University of Science and Technology (NTNU), Norway}
\author{Dirk Gillespie}%
 \email{dirk\_gillespie@rush.edu}
\affiliation{Department of Biophysics and Physiology, Rush University Medical Center, Chicago, USA}

\date{\today}

\begin{abstract}
The Mean Spherical Approximation (MSA) is a commonly-used thermodynamic theory for computing the energetics of ions in the primitive model (i.e., charged hard-sphere ions in a background dielectric). For the excess chemical potential, however, the early MSA formulations (which were widely adopted) only included the terms needed to compute the mean excess chemical potential (or the mean activity coefficient). Other terms for the chemical potential $\mu_i$ of individual species $i$ were not included because they sum to $0$ in the mean chemical potential. Here, we derive these terms to give a complete MSA formulation of the chemical potential. The result is a simple additive term for $\mu_i$ that we show is a qualitative improvement over the previous MSA version. In addition, our derivation shows that the MSA's assumption of global charge neutrality is not strictly necessary, so that the MSA is also valid for systems close to neutrality. 
\end{abstract}

\begin{quotation}
\noindent
\centering
\textbf{The following article has been submitted to  \emph{The Journal of Chemical Physics}.\\
After it is published, it will be found at\\
https://aip.scitation.org/journal/jcp.}
\end{quotation}
\maketitle

\section{Introduction}
The Mean Spherical Approximation (MSA) \cite{blum75,blum80} has long been used to compute the energetics of model electrolyte solutions like the primitive model, where ions are charged hard spheres in a background dielectric. While it has thermodynamic inconsistencies,\cite{hoye77b} the power the MSA has been its simple analytic formulas for thermodynamic properties like the electrostatic contributions to the Helmholtz, Gibbs, and internal energies, as well as the entropy and excess chemical potentials.

A recent study \cite{gillespie21} compared the MSA’s excess chemical potential for the individual ion species (denoted ${{\mu }_{i}}$ for species $i$) to Monte Carlo (MC) simulations and found that the MSA deviated significantly at high electrolyte concentrations. In that paper, Gillespie et al. \cite{gillespie21} derived a new theory of the electrostatic excess chemical potential which corrected this deviation. The needed correction was adding a new term to an MSA-like excess chemical potential that was proportional to ${{z}_{i}}$, the valence of ion species $i$. In the commonly cited and commonly used literature of the MSA (reviewed, for example, in Ref.~\onlinecite{nonner00}), terms proportional to ${{z}_{i}}$ do not appear in $\mu_i$.

The focus of this work is terms of this type within the MSA. In early MSA papers such terms are mentioned (e.g., \onlinecite{hoye78}), but they were not explicitly written out. Rather, they were commonly lumped into a “$+{{z}_{i}}\cdot \text{const}\text{.}$” term \cite{hoye78,hoye21} because, by charge neutrality, such terms do not contribute to the mean excess chemical potential (or mean activity coefficient) of the electrolyte. Thus, the formulas given for $\mu_i$ are incomplete. Here, we derive the missing terms.

Having complete formulas for ${{\mu }_{i}}$ is important because often the incomplete formulas are used to study the energetics of individual ion species. For example, any calculation of ions in equilibrium (e.g., inhomogeneous electrolytes in an electrical double layer) requires, by the definition of chemical equilibrium in the grand canonical ensemble, an accurate value for ${{\mu }_{i}}$ for each ion species individually. Moreover, many theories apply the MSA formulas of a homogeneous system repeatedly to compute ion densities in inhomogeneous systems (e.g., classic density functional theory \cite{rosenfeld93,gillespie02,roth16}).

Our new terms that are proportional to ${{z}_{i}}$ are derived from the early MSA papers that list the starting points, but not the final formulations. Thus, we are building onto previous well-established results, adding back some lost pieces. When this extended MSA is compared to MC simulations, the result is a significant improvement in the excess chemical potentials.

This work is laid out as follows. First, in Sec.\ \ref{notneutral} we show how Blum's original derivation of the MSA \cite{blum75,blum80} is still valid close to, but away from, charge neutrality. This is necessary because calculating individual species chemical potentials involves adding a small amount of unbalanced charge to a charge neutral system. Next, in Sec.\ \ref{internalene}, we lay an additional foundation with a discussion of the MSA internal energy. In Sec.\ \ref{chempotls}, we derive our main result for the individual species electrostatic excess chemical potentials (Eq.\ (\ref{mui})). Lastly, in Sec.\ \ref{Results} we compare the previous and new formulations to MC simulations to show the qualitative improvement of our extended MSA chemical potential compared over the standard MSA version.

\section{Theory}
Throughout we generally employ the notation and results of Ref.~\onlinecite{hoye21}.

\subsection{The MSA for electrolytes close to neutrality}
\label{notneutral}

The MSA solution obtained by Blum for an asymmetric electrolyte is for a system with strict with charge neutrality imposed.\cite{blum75} However, to compute individual species chemical potentials, a single ion is added to the system and charge neutrality is violated. Here, we show that the usual MSA solution is still valid for systems close to neutrality. This also validates the use of MSA formulas in non-uniform system with local deviations from charge neutrality (e.g., Refs.~\onlinecite{gillespie02,roth16}).

We start with Blum's solution \cite{blum75} to the Baxter factorization functions $Q_{ij}(r)$ of ion species $i$ and $j$ that define the direct correlation functions $c_{ij}(r)$ by
\begin{equation}
    \delta_{ij}-(\rho_i \rho_j)^{1/2} \tilde c_{ij}(k)=\sum_l \tilde Q_{il}(k) \tilde Q_{jl}(k)
    \label{OZ}
\end{equation}
where the tilde denotes the Fourier transform and $\rho_i$ is the density of species $i$.
Finding the $Q_{ij}(r)$ is possible for a system of Yukawa interaction potentials
\begin{equation}
    \beta \psi_{ij}(r,Z)=\frac{\beta e^2 z_i z_j}{4\pi \varepsilon \varepsilon_0}\frac{e^{-Zr}}{r}.
\end{equation}
Here, $\beta =1/k_BT$ (with $k_B$ being the Boltzmann constant and $T$ the absolute temperature), $e$ is the fundamental charge, $\varepsilon $ is the dielectric constant of the system, and ${{\varepsilon }_{0}}$ is the permittivity of free space.
Coulomb interactions are the limit $Z\rightarrow0$, a limit that will be important throughout.

The MSA enforces the boundary condition
\begin{equation}
\label{cijbc}
    c_{ij}(r) = -\beta \psi_{ij}(r,Z) \quad (r>\sigma_{ij})
\end{equation}
where $\sigma_{ij}=(\sigma_i+\sigma_j)/2$ with $\sigma_i$ the hard core diameter of species $i$. Therefore, the $c_{ij}(r)$ always have the form
\begin{equation}
c_{ij}(r)=
\begin{cases} 
c_{Iij}(r) & r < \sigma_{ij} \\
-\beta\psi_{ij}(r,Z) & r > \sigma_{ij}
\end{cases}
\label{85}
\end{equation}
with the unknown function $c_{Iij}(r)$ when the hard cores of the ions overlap.
The other unknown correlation function $h_{ij}(r)$ is related to $c_{ij}(r)$ via the Ornstein-Zernike equation
\begin{equation}
\label{OZreal}
    h_{ij}(r) = c_{ij}(r) + \sum_l \rho_l \int c_{il}(r) h_{lj}(|\mathbf{r}-\mathbf{r'}|) d \mathbf{r'}
\end{equation}
and its Fourier transform
\begin{equation}
\label{OZch}
    \tilde h_{ij}(k) = \tilde c_{ij}(k) + \sum_l \rho_l \tilde c_{il}(k) \tilde h_{lj}(k).
\end{equation}
It has boundary conditions
\begin{equation}
\label{hijbc}
    h_{ij}(r) = -1 \quad (r<\sigma_{ij}).
\end{equation}
This total correlation function is related to the pair correlation function $g_{ij}(r)$ via
\begin{equation}
\label{gij}
    g_{ij}(r) = h_{ij}(r) + 1.
\end{equation}

The Baxter factor functions are
\begin{equation}
\label{Qij}
    Q_{ij}(r)=
    \begin{cases} 
 q_{ij}(r)-z_i a_j e^{-Zr} & r < \sigma_{ij} \\
-z_i a_j e^{-Zr} & r > \sigma_{ij}
\end{cases}
\end{equation}
where $a_j$ is defined later in Eq.\ (\ref{ai}). 
To find $q_{ij}(r)$, we start with Blum's derivation \cite{blum75} one step before an invocation of charge neutrality, namely at his Eq.\ (2.23):
\begin{eqnarray}
\label{223}
    J_{ij}(r) & = & Q_{ij}(r) + \sum_k \rho_k \int_{\lambda_{jk}}^{\sigma_{jk}} Q_{kj}(r') J_{ik}(|r-r'|) dr'\\
    & & -e^{-Zr} \sum_k \rho_k \int_{{\sigma_{jk}-r}}^\infty J_{ik}(r')A_{kj} e^{-Zr'} dr'
    \nonumber
\end{eqnarray}
where $\lambda_{jk} = (\sigma_j - \sigma_k)/2$,
\begin{equation}
    J_{ij}(r) = 2\pi \int_r^\infty t h_{ij}(t) dt
\label{Jij}
\end{equation}
and
\begin{equation}
    A_{ij} = z_i a_j.
\end{equation}
The next step in Blum's solution is to invoke the condition that the ion cloud around a central ion neutralizes that ion's charge:
\begin{equation}
\label{gijneutral}
     -z_i = 4\pi \sum_k z_k \rho_k \int_0^\infty r^2 h_{ik}(r) dr.
\end{equation}
In the MSA, this relation about \emph{local} charge neutrality around a central ion is true even without \emph{global} charge neutrality (i.e., $\sum_k z_k \rho_k = 0$). This was shown by Blum, where he derived Eq.\ (\ref{gijneutral}) as Eq.\ $(4.18)$ of Ref.~\onlinecite{blum80}. However, in the original MSA derivation, Blum \cite{blum75} used Eq.\ (\ref{gijneutral}) with $h_{ik}(r) \mapsto g_{ik}(r)$, which adds an unnecessary use of global charge neutrality.

Next, using integration by parts, we have
\begin{equation}
    4\pi \sum_k z_k \rho_k \int_0^\infty r^2  h_{ik}(r) dr =  2 \sum_k z_k \rho_k \int_0^\infty J_{ik}(r)dr
\end{equation}
so that, by Eq.\ (\ref{gijneutral}),
\begin{equation}
    \sum_k \rho_k \int_0^\infty J_{ik}(r)A_{kj}dr = -\frac{1}{2}A_{ij}.
\end{equation}
This is Blum's Eq.\ (2.25) of Ref.~\onlinecite{blum75}, and so we can take the limit $Z\rightarrow0$ just as he did to get Blum's Eq.\ (2.26):
\begin{eqnarray}
\label{226}
    J_{ij}(r) & = & Q_{ij}(r) + \frac{1}{2}A_{ij}+\frac{1}{2}a_j \delta q \\
    \nonumber
    & & + \sum_k \rho_k \int_{\lambda_{jk}}^{\sigma_{jk}} Q_{kj}(r') J_{ik}(|r-r'|) dr'\\
    & & + \sum_k \rho_k \int^{{\sigma_{jk}-r}}_0 J_{ik}(r')A_{kj}  dr'.
    \nonumber
\end{eqnarray}
This equation is the same as Blum's, but without invoking global charge neutrality.  However, charge neutrality was invoked once more in his analysis.

Blum next notes that, by Eqs.\ (\ref{hijbc}) and (\ref{Jij}), for $r<\sigma_{ij}$,
\begin{equation}
    J_{ij}(r) = \pi r^2 + J_{ij}(0)
\end{equation}
so that when $r<\sigma_{ij}$
\begin{eqnarray}
\label{226x}
    \pi r^2 + J_{ij}(0) & = & Q_{ij}(r) + \frac{1}{2}A_{ij}\\
    \nonumber
    & & + \sum_k \rho_k \int_{\lambda_{jk}}^{\sigma_{jk}} Q_{kj}(r') \left( \pi (r-r')^2 + J_{ik}(0) \right) dr'\\
    & &  \pi \sum_k \rho_k A_{kj} \int^{{\sigma_{jk}-r}}_0 \left(  r'^2 + J_{ik}(0) \right) dr'.
    \nonumber
\end{eqnarray}
The place where charge neutrality is invoked is in the evaluation of the last integral:
\begin{eqnarray}
\nonumber
& & \sum_k \rho_k A_{kj} \int^{{\sigma_{jk}-r}}_0 \left(  r'^2 + J_{ik}(0) \right) dr'\\
&=&
\sum_k \rho_k A_{kj} J_{ik}(0) (\sigma_{jk}-r) \\
\nonumber
& &
+\frac{a_j}{3} \sum_k z_k \rho_k \left(  \sigma_{jk}-r \right)^3.
\end{eqnarray}
Expanding the cubic gives a new term for Eq.\ (\ref{226x}), namely
\begin{equation}
    -r^3 \frac{\pi}{3} a_j \sum_k z_k \rho_k,
\end{equation}
plus corresponding induced changes in the $r^n$ ($n=0,1,2$) terms.
This new $r^3$ term is asymptotically small as charge neutrality is approached.

Blum's analysis continues (described in more detail in Ref.~\onlinecite{blum80} than in Ref.~\onlinecite{blum75}) by using Eq.\ (\ref{226x}) to show that (with that one invocation of charge neutrality) the $q_{ij}(r)$ in Eq.\ (\ref{Qij}) (now denoted $q_{ij}^{\textrm{MSA}}(r))$ are quadratic in $r$:
\begin{equation}
\label{qij}
    q_{ij}^{\textrm{MSA}}(r)=(r-\sigma_{ij})q'_{ij}+\frac{1}{2}(r-\sigma_{ij})^2q''_j.
\end{equation}
The quantities $q'_{ij}$ and $q''_j$ are defined later in Eqs.\ (\ref{qp}) and (\ref{qpp}).
Here, we see that with a small violation of charge neutrality the functional forms of $q_{ij}(r)$ (and therefore the $Q_{ij}(r,Z)$) are altered only infinitesimally:
\begin{equation}
\label{newqij}
    q_{ij}(r)=q_{ij}^{\textrm{MSA}}(r)  - r^3 \frac{\pi}{3} a_{j} \sum_k z_k \rho_k.
\end{equation}

By Eq.\ (\ref{OZ}), these small new terms in $Q_{ij}(r,Z)$ produce a perturbation in the $c_{ij}(r)$, which we denote $\delta c_{ij}(r)$.
By Eqs.\ (\ref{cijbc}) and (\ref{85}),
\begin{equation}
\delta c_{ij}(r)=
\begin{cases} 
\delta c_{Iij}(r) & r < \sigma_{ij} \\
0 & r > \sigma_{ij}
\end{cases}
\label{90}
\end{equation}
because for $r>\sigma_{ij}$ the $c_{ij}(r)$ remain unchanged, fixed by the MSA ansatz defined by Eq.\ (\ref{cijbc}). Next, we show that this perturbation does not significantly alter the excess Helmholtz free energy density $A$ of the system.

It is possible to obtain an expression for $A$ in terms of the $c_{ij}(r)$ via the quantity $I=-\beta A$. Specifically, H\o ye and Stell \cite{hoye77b} showed that
\begin{eqnarray}
\label{84}
I&=&\frac{1}{2}\sum_{i,j}\rho_i\rho_j(\tilde c_{ij}(0)-\tilde c_{ij}^{\textrm{HS}}(0))\\
\nonumber
& &-\frac{1}{2}\frac{1}{(2\pi)^3}\textrm{Trace}\int\{\ln[1-\rho \tilde c(k)]+\rho \tilde c(k)\}\,d{\bf k}\\
& &+\frac{1}{2}\frac{1}{(2\pi)^3}\textrm{Trace}\int\{\ln[1-\rho \tilde c^{\textrm{HS}}(k)]+\rho \tilde c^{\textrm{HS}}(k)\}\,d{\bf k},
\nonumber
\end{eqnarray}
where $c(r)$ are matrices with matrix elements $c_{ij}(r)$ and $\rho$ is the vector of densities $\rho_i$. In this equation and throughout, the superscript $\textrm{HS}$ denotes the reference system of hard spheres.
By substituting $c_{ij}(r) \mapsto c_{ij}(r) + \delta c_{ij}(r)$ and using Eq.\ (\ref{OZch}), one finds that the leading order error in $I$ is
\begin{eqnarray}
\nonumber
\delta I&=&\frac{1}{2}\sum_{i,j}\rho_i\rho_j\,\delta\tilde c_{ij}(0)\\
\nonumber
&&+\frac{1}{2}\frac{1}{(2\pi)^3}\textrm{Trace}\int\frac{\rho\tilde c(k)}{1-\rho\tilde c(k)}\rho\,\delta\tilde c(k)\,d{\bf k}\\
\nonumber
&=&\frac{1}{2}\sum_{i,j}\rho_i\rho_j\,\delta\tilde c_{ij}(0)+\frac{1}{2}\frac{1}{(2\pi)^3}\textrm{Trace}\int \tilde h(k)\rho\,\delta\tilde c(k)\,d{\bf k}\\
&=&\frac{1}{2}\sum_{i,j}\rho_i\rho_j\,\delta\tilde c_{ij}(0)+\frac{1}{2}\textrm{Trace}\int \rho h(r) \rho\,\delta c(r)\,d{\bf r}.
\label{91}
\end{eqnarray}
For the last integral in expression (\ref{91}), because $h_{ij}(r)=-1$ for $r<\sigma_{ij}$ and $\delta c_{ij}(r)=0$ for $r>\sigma_{ij}$, we have
\begin{eqnarray}
\nonumber
&&\frac{1}{2}\textrm{Trace}\int \rho h(r) \rho\,\delta c(r)\,d{\bm{r}}\\
&=&-\frac{1}{2}\textrm{Trace}\,\rho \rho\int_{r<\sigma_{ij}} \delta c(r)\,d{\bm{r}}\\
&=&-\frac{1}{2}\sum_{i,j}\rho_i\rho_j\,\delta\tilde c_{ij}(0)
\label{}
\end{eqnarray}
where in the last step we used the generic identity
\begin{equation}
\label{ftildeidentity}
    \tilde f(0) = \int f(r)\,d{\bf r}.
\end{equation} 
Thus,
\begin{equation}
\delta I = 0.
\label{92}
\end{equation}

From this it follows that any change to $A$ is second-order for small errors like $\delta c_{Iij}(r) \sim r^3 \sum_k z_k \rho_k$. Consequently, the chemical potentials 
\begin{equation}
\label{partialA}
    \mu_i = \frac{\partial A}{\partial \rho_i}
\end{equation}
are the same as in the charge neutral case because, it being a derivative, any violation of charge neutrality is infinitesimal. (Appendix \ref{Appneuback} shows that this is also true for a different but related case, where there is constant neutralizing background.) Similarly, the excess internal free energy density $E$ is only perturbed to second-order since $E=\partial \beta A / \partial \beta$.

Thus, we conclude that to leading order the MSA equations for the neutral electrolyte are also be valid away from neutrality, but close to it. Therefore, in the following we can use the standard MSA equations to derive formulas for the individual species' chemical potential.

\subsection{MSA excess internal energy per particle}
\label{internalene}

An important quantity we will need is $u_i$, the excess internal energy per particle of species $i$. Comparison of different formulations of this quantity away from charge neutrality requires a little care.

H\o ye and Blum \cite{hoye78} stated that
\begin{equation}
\label{ui}
    	\beta {{u}_{i}}=\frac{{{\alpha }^{2}}}{4\pi }{{z}_{i}}{{N}_{i}}+{{z}_{i}}\beta {{u}^{*}}
\end{equation}
where ${{\alpha }^{2}}=\beta {{e}^{2}}/\varepsilon {{\varepsilon }_{0}}$,
\begin{equation}
\label{Ni}
    	{{N}_{i}}=-\frac{\Gamma {{z}_{i}}+\eta {{\sigma }_{i}}}{1+\Gamma {{\sigma }_{i}}},
\end{equation}
and
\begin{equation}
\label{ustar}
    	\beta {{u}^{*}}=-\frac{{{\alpha }^{2}}}{24 }\sum\limits_{l}{{{\rho }_{l}}\sigma _{l}^{2}\left( {{N}_{l}}{{\sigma }_{l}}+\frac{3}{2}{{z}_{l}} \right)}.
\end{equation}
$\Gamma $ is the MSA screening length parameter given implicitly by
\begin{equation}
\label{Gamma}
    	4{{\Gamma }^{2}}={{\alpha }^{2}}\sum\limits_{i}{{{\rho }_{i}}{{\left( \frac{{{z}_{i}}-\eta \sigma _{i}^{2}}{1+\Gamma {{\sigma }_{i}}} \right)}^{2}}}
\end{equation}
with $\eta =L{{P}_{n}}/2$ for $L=\pi /\Delta $, 
\begin{equation}
\label{Delta}
   	\Delta =1-\frac{\pi }{6}{{\zeta }_{3}}
\end{equation}
\begin{equation}
    	{{\zeta }_{n}}=\sum\limits_{k}{{{\rho }_{k}}\sigma _{k}^{n}}
\end{equation}
\begin{equation}
\label{Pn}
    	{{P}_{n}}=\frac{1}{\Omega }\sum\limits_{k}{\frac{{{z}_{k}}{{\rho }_{k}}{{\sigma }_{k}}}{1+\Gamma {{\sigma }_{k}}}}
\end{equation}
\begin{equation}
\label{Om}
    	\Omega =1+\frac{L}{2}\sum\limits_{k}{\frac{{{\rho }_{k}}\sigma _{k}^{3}}{1+\Gamma {{\sigma }_{k}}}}.
\end{equation}
It is the $z_i \beta u^*$ in Eq.\ (\ref{ui}) that has generally been ignored and written off as “$+{{z}_{i}}\cdot \text{const}\text{.}$” for subsequent derivations of $\mu_i$.

A derivation of Eq.\ (\ref{ui}) is given in Appendix \ref{Appustar}, as H\o ye and Blum \cite{hoye78} merely stated this non-trivial result. 
An important aspect of this derivation is that this $u_i$ is the excess internal energy beyond the mean-field internal energy $u_i^\textrm{MF}$.
Specifically,
\begin{equation}
    \beta u_i = \frac{1}{2} \sum_j \rho_j \int h_{ij}(r) \beta \psi_{ij}(r,Z) d \mathbf{r}
    \label{uidef}
\end{equation}
and
\begin{eqnarray}
    \beta u_i^\textrm{MF} &=& \frac{1}{2} \sum_j \rho_j \int \beta \psi_{ij}(r,Z) d \mathbf{r}\\
    &=&\frac{1}{2} \sum_j \rho_j \int_{r<\sigma_{ij}} (\beta \psi_{ij}(r,Z) + c_{ij}(r)) d \mathbf{r}\\
    \nonumber
    &&-\frac{1}{2} \sum_j \rho_j \tilde c_{ij}(0).
\end{eqnarray}
where the last term follows from Eqs.\ (\ref{cijbc}) and  (\ref{ftildeidentity}). While $u_i$ is always finite, for a non-neutral Coulombic system $u_i^\textrm{MF} = \infty$ because $\tilde c_{ij}(0) = \infty$; in a neutral system, $u_i^\textrm{MF} = 0$. The sum $u_i + u_i^\textrm{MF}$ is the full internal energy per particle.

This distinction is important because, as shown by H\o ye and Stell, \cite{hoye77b} the Ornstein-Zernike equation (Eq.\ (\ref{OZreal})) and its MSA boundary conditions Eqs.\ (\ref{cijbc}) and (\ref{hijbc}) allow for a simple statement of the full internal energy via Eq.\ (\ref{gij}):
\begin{eqnarray}
\label{internalenedef}
\beta u_i + \beta u_i^\textrm{MF} & \equiv & \frac{1}{2} \sum_j \rho_j \int g_{ij}(r) \psi_{ij}(r,Z) d \mathbf{r}\\
& = & -\frac{1}{2} \sum_j \rho_j \int (h_{ij}(r) + 1) c_{ij}(r) d \mathbf{r}\\
& = & \frac{1}{2} \left( c_{ii}(0) + 1 \right) - \frac{1}{2} \sum_j \rho_j \tilde c_{ij}(0).
\label{hoyeinternalene}
\end{eqnarray}
From this formulation, H\o ye and Stell \cite{hoye77b} derived the full Helmholtz free energy density (i.e., not subtracting off the Helmholtz of the mean-field contribution). Differentiating that, they found the chemical potential with the mean-field term $\mu_i^\textrm{MF}$ included:
\begin{eqnarray}
\nonumber
&&\beta \mu_i + \beta \mu_i^\textrm{MF}\\
& = &
\label{ctildemethod2}
\beta u_i + \beta u_i^\textrm{MF} + \frac{1}{2} \sum_k \rho_k \left( \tilde c_{ik}^{\textrm{HS}}(0) - \tilde c_{ik}(0) \right)\\
&=&2(\beta u_i + \beta u_i^\textrm{MF} ) + \frac{1}{2} \left( c_{ii}^{\textrm{HS}}(0) - c_{ii}(0) \right).
\label{cmethod2}
\end{eqnarray}
Eq.\ (\ref{cmethod2}) follows from Eq.\ (\ref{ctildemethod2}) by applying Eq.\ (\ref{hoyeinternalene}). \cite{hoye21}

Before continuing on to the individual species excess chemical potentials, we note that in Sec.\ \ref{notneutral} the derivation of Eq.\ (\ref{92}) includes all mean-field terms and those terms technically
diverge due to the divergence of $\tilde c_{ij}(0)$ in the $Z\rightarrow0$ limit. However, the integral over the Coulomb interactions (for $0<r<\infty$) can be separated out as mean-field terms, and it is easily seen that that they have no influence upon $\delta I$ and vice versa. This also means that mean-field terms do not interfere with the solution of the Ornstein-Zernike equation since they are not involved in correlations.

\subsection{MSA individual species chemical potentials}
\label{chempotls}

The ion species' excess chemical potential beyond the mean-field can be obtained from either Eq.\ (\ref{ctildemethod2}) or Eq.\ (\ref{cmethod2}) by subtracting off the mean-field contribution. From basic mean field theory we know that
\begin{equation}
    \mu_i^\textrm{MF} = 2u_i^\textrm{MF}.
\end{equation}
It is then clear that Eq.\ (\ref{cmethod2}) is more convenient to use because subtracting off $2u_i^\textrm{MF}$ is straightforward:
\begin{equation}
\label{cmethod}
   \beta \mu_i = 2\beta u_i + \frac{1}{2} \left( c_{ii}^{\textrm{HS}}(0) - c_{ii}(0) \right).
\end{equation}
All these terms are finite for both neutral and non-neutral systems as $Z\rightarrow0$. \cite{hoye21}

Using Eq.\ (\ref{ctildemethod2}), on the other hand, is more complicated, even though it is equivalent to Eq.\ (\ref{cmethod2}). That is because there is no convenient formulation for 
\begin{equation}
    \frac{1}{2} \sum_k \rho_k \tilde c_{ik}(0) - \beta u_i^\textrm{MF},
\end{equation}
and therefore one cannot easily determine what remains after the infinities cancel. (See Ref.~\onlinecite{hoye21} for a discussion of $Z^{-1}$ terms in $\tilde c_{ik}(0)$ that lead to divergences without charge neutrality when $Z\rightarrow0$.)

Therefore, we use Eq.\ (\ref{cmethod}) to derive “$+{{z}_{i}}\cdot \text{const}\text{.}$” terms. We start with \cite{hoye21}
\begin{eqnarray}
\label{deltacii}
\nonumber
    2\pi \left( c_{ii}^{\text{HS}}(0)-{{c}_{ii}}(0) \right)&=&{q''_{i}}-q_i''^{\textrm{HS}}-{{z}_{i}}\sum\limits_{k}{{{\rho }_{k}}{{a}_{k}}{q'_{ik}}} \\ 
 & &+\sum\limits_{k}{{{\rho }_{k}}\left( \kappa _{ik}^{\text{HS}} q_k''^{\textrm{HS}}-{{\kappa }_{ik}}q_k'' \right)}
\end{eqnarray}
where
\begin{eqnarray}
\nonumber
\kappa_{ij}&\equiv&\int_{\lambda_{ji}}^{\sigma_{ij}} q_{ij}(r) dr\\
&=&-\frac{1}{2}q_{ij}' \sigma_i^2+\frac{1}{6}q_{j}'' \sigma_i^3
+\delta \kappa_{ij} ,
\label{kappaij}
\end{eqnarray}
\begin{equation}
\label{qp}
    q_{ij}' = L\left(\sigma_i+\sigma_j+\frac{L}{2}\zeta_2\sigma_i\sigma_j\right)-\frac{\Gamma^2}{\alpha^2}a_ia_j,
\end{equation}
\begin{equation}
\label{qpp}
 q_j'' = 2L\left(1+\frac{L}{2}\zeta_2\sigma_j\right)+LP_na_j,
\end{equation}
\begin{equation}
    \zeta_n = \sum_k(\sigma_k)^n,
\end{equation}
\begin{equation}
   q_k''^{\textrm{HS}}=2L\left( 1+\frac{L}{2}{{\zeta }_{2}}{{\sigma }_{j}} \right),
\end{equation}
and
\begin{equation}
\label{ai}
    a_i = \alpha^2\frac{z_i-\eta\sigma_i^2}{2\Gamma(1+\Gamma\sigma_i)}.
\end{equation}
The small correction to $\kappa_{ij}$ from non-charge-neutrality is
\begin{equation}
   \delta \kappa_{ij} = - \frac{\pi}{12}a_j \left( \sigma_{ij}^4-\lambda_{ij}^4 \right)  \sum_k z_k \rho_k ,
\end{equation}
plus similar contributions from changes in $r^n$ ($n=0,1,2$) terms.
Note that these terms contain only infinitesimally small $\delta \kappa_{ij}$ corrections to standard MSA formulations. In the following we will drop these, as they are arbitrarily small.

To derive the “$+{{z}_{i}}\cdot \text{const}\text{.}$” terms, we evaluate Eq.\ (\ref{deltacii}) term by term. For the first terms in Eq.\ (\ref{deltacii}), by Eq.\ (A20) in Appendix 2 of Ref.~\onlinecite{hoye21}, we have
\begin{equation}
    {q''_{i}}-q_{i}''^{\textrm{HS}}=\delta q''_{i}+{{z}_{i}}\delta q^*
\label{ziterm}
\end{equation}
with
\begin{equation}
	\delta {q''_{i}}=-\pi \frac{{{P}_{n}}}{\Delta }{{a}_{i}}\Gamma {{\sigma }_{i}}-{{\pi }^{2}}{{\left( \frac{{{P}_{n}}}{\Delta } \right)}^{2}}\frac{{{\alpha }^{2}}}{\Gamma }\sigma _{i}^{2}
\end{equation}
and
\begin{equation}
\label{deltaqstar1}
    \delta {q^{*}} \equiv \frac{\pi }{2}\frac{{{P}_{n}}}{\Delta }\frac{{{\alpha }^{2}}}{\Gamma }.
\end{equation}

For the next term in Eq.\ (\ref{deltacii}), by Eq.\ (\ref{qp}) we have
\begin{eqnarray}
\label{sum}
   \sum\limits_{k}{{{\rho }_{k}}{{a}_{k}}q'_{ik}}&=&
   L{{\sigma }_{i}}\sum\limits_{k}{{{\rho }_{k}}{{a}_{k}}}+\frac{{{L}^{2}}}{2}{{\zeta }_{2}}{{\sigma }_{i}}\sum\limits_{k}{{{\rho }_{k}}{{\sigma }_{k}}{{a}_{k}}}\\ 
 && - \frac{2{{\Gamma }^{2}}}{{{\alpha }^{2}}}{{a}_{i}}\sum\limits_{k}{{{\rho }_{k}}a_{k}^{2}}+L\sum\limits_{k}{{{\rho }_{k}}{{\sigma }_{k}}{{a}_{k}}}.  
\nonumber
 \end{eqnarray}
We now note that the first three terms all have coefficients ${{\sigma }_{i}}$ or ${{a}_{i}}$ before the sum. Thus, they cannot contribute any terms of the form ${{z}_{i}}\cdot \text{const}\text{.}$ where the constant is independent of $i$; only the last term can contribute such a term. This last term was evaluated in Eq.\ (A4) of Appendix 1 of Ref.~\onlinecite{hoye21}, with the result
\begin{equation}
\label{deltaqstar2}
    	L\sum\limits_{k}{{{\rho }_{k}}{{\sigma }_{k}}{{a}_{k}}}=L\frac{{{\alpha }^{2}}}{2\Gamma }{{P}_{n}} = \delta {q^{*}}.
\end{equation}
More details for evaluating Eq.\ (\ref{sum}) are given in Appendix \ref{secA2}.

For the last term in Eq.\ (\ref{deltacii}), we use the definition of the ${{\kappa }_{ij}}$ in Eq.\ (\ref{kappaij}) to see that all summands in an expansion of this term will have a power of ${{\sigma }_{i}}$ as a coefficient and therefore cannot lead to terms of the form ${{z}_{i}}\cdot \text{const}$. Therefore, the third term in Eq.\ (\ref{deltacii}) contributes only to the previously derived MSA chemical potential.

Combining these results, we have that the two $\delta {q^{*}}$ terms in Eqs.\ (\ref{deltaqstar1}) and (\ref{deltaqstar2}) cancel; only $2z_i\beta u_i^*$ from the $2\beta u_i$ term in Eq.\ (\ref{cmethod}) remains beyond the standard MSA excess chemical potential formulation. Therefore,
\begin{equation}
\label{mui}
    	{{\mu }_{i}}=\mu _{i}^{\text{MSA}}+2{{z}_{i}}{{u}^{*}}
\end{equation}
where $u^*$ is given by Eq.\ (\ref{ustar}) and $\mu _{i}^{\text{MSA}}$ is the previous MSA excess chemical potential given by \cite{hoye21}
\begin{equation}
\label{muiMSA}
    \beta \mu _{i}^{\text{MSA}}=-\frac{{{\alpha }^{2}}}{4\pi }\left[ \frac{z_{i}^{2}\Gamma }{1+\Gamma {{\sigma }_{i}}}+\eta {{\sigma }_{i}}\left( \frac{2{{z}_{i}}-\eta \sigma _{i}^{2}}{1+\Gamma {{\sigma }_{i}}}+\frac{\eta \sigma _{i}^{2}}{3} \right) \right].
\end{equation}

We note that $u^*=0$ when all the ions have the same size (i.e., $\sigma_i=\sigma$ for all $i$), so our results are identical to the standard MSA results in the restricted primitive model. Moreover, nothing about the MSA quantities like $\Gamma$ are changed; their formulas are the same as previous standard MSA formulas since we are only restoring chemical potential terms that were dropped when only the mean chemical potential (activity coefficient) were sought.

To derive Eq.\ (\ref{mui}), we used the existing MSA formulation, but where we considered that there was a small (infinitesimal) violation of charge neutrality so that it is valid to take the derivative in Eq.\ (\ref{partialA}). Appendix \ref{Appneuback} derives Eq.\ (\ref{mui}) for a different but related system, where there is a constant neutralizing background charge density, instead of a violation of charge neutrality. Thus, our main result (Eq.\ (\ref{mui})) can be obtained in two very different ways. In Sec.\ \ref{Results}, we will show that it agrees very well with simulation results for $\mu_i$.

In Appendix \ref{secA2}, Eq.\ (\ref{sum}) is verified in more detail. Specifically, with $a_i$ given by Eq.\ (\ref{ai}), it is possible to separate out a constant from its $x^2/(1+x)=x-1+1/(1+x)$ part, where $x=\Gamma\sigma_i$. This could potentially change the $+{{z}_{i}}\cdot \text{const}\text{.}$ term in $\mu_i$ calculated from Eq.\ (\ref{sum}) (and potentially in $\mu_i^{\textrm{MSA}}$ as well).
That this uncertainty about $a_i$ does not change result (\ref{mui}) is shown in Appendix \ref{secA2}. While we do not show it, results from Ref.~\onlinecite{hoye21} can be used to show the same for $\mu_i^{\textrm{MSA}}$.

\begin{figure*}[htp]
    \centering
    \includegraphics[width=6.5in]{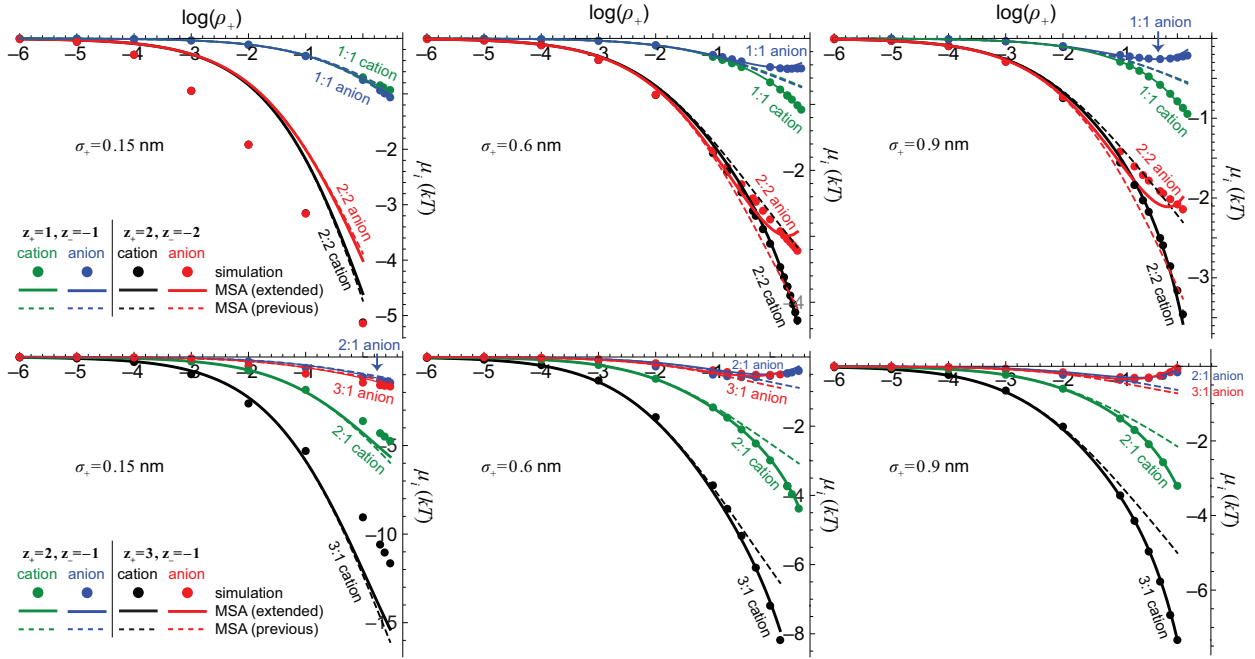}
    \caption{Comparison of our extended MSA $\mu_i$ from Eq.\ (\ref{mui}) (solid lines) to both the standard MSA formulation Eq.\ (\ref{muiMSA}) (dashed lines) and MC simulations (symbols). Each panel shows the results for $\mu_i$ versus $\rho_+$ for two electrolytes with different ion valences $z_+$ and $z_-$, but the same cation sizes. One electrolyte is depicted with cations and anions in green and blue colors, respectively, and the other electrolyte is depicted with cations and anions in black and red colors, respectively. The dielectric constant values is always $\varepsilon=78.45$ and $T=298.15$ K. The anion diameter is always $0.3$ nm, while the cation diameter $\sigma_+$ is indicated in each panel. 
    }
    \label{fig:compare1}
\end{figure*}

\section{Results and discussion}
\label{Results}

We verify Eq.\ (\ref{mui}) by comparing it to the chemical potentials from grand canonical MC simulations of homogeneous fluids of charged, hard spheres. This is then a direct comparison between identical systems.

The details of the simulations are given in Ref.~\onlinecite{gillespie21}. To summarize, each ion species’ total chemical potential $m_i$ is the sum of ideal gas, hard-sphere, and electrostatic components:
\begin{equation}
    \beta m_i = \ln \rho_i + \beta \mu_i^\textrm{HS} + \beta \mu_i.
\end{equation}
The last term ($\mu_i$) is what the MSA computes and is our focus. In MC simulations, the total chemical potential $m_i$ for each species must be specified, not the densities $\rho_i$. In the MSA, the reverse is true. Therefore, in the simulations, values for the $m_i$ are iterated until the desired ion concentrations ${{\rho }_{i}}$ are achieved, using a well-established algorithm. \cite{malasics10} Since now both the $m_i$ and the $\rho_i$ are now known, so is $\beta m_i - \ln \rho_i = \beta \mu_i^\textrm{HS} + \beta \mu_i$. Thus, once the hard-sphere component is determined, so is the desired $\mu_i$ (by subtraction). The hard-sphere component $\mu_i^\textrm{HS}$ is computed from the acceptance/rejection ratio of attempts to insert uncharged hard sphere “ions,” as previously described. \cite{boda11} This process works until, at high enough ion packing fractions, the correlation lengths between ions become too long to be contained in reasonably sized simulation box.

The first comparisons are shown in Fig.\ \ref{fig:compare1}. Our extended MSA $\mu_i$ from Eq.\ (\ref{mui}) (solid lines) gives significantly better values for both $\mu_+$ and $\mu_-$ compared to $\mu_+^{\textrm{MSA}}$ and $\mu_-^{\textrm{MSA}}$ from Eq.\ (\ref{muiMSA}) (dashed lines). In fact, for many cases the results are qualitatively better. For example, for monovalent anions ($z_i=-1$), $\mu_-$ has a minimum in both the simulations and our formulation, while the MSA is always monotonic for both ions. 

Also, for $2:2$ electrolytes with large ions, both $\mu_+^{\textrm{MSA}}$ and $\mu_-^{\textrm{MSA}}$ are qualitatively incorrect; the cation curves (black dashed lines in the top row of Fig.~\ref{fig:compare1}) follow the anion simulation results (red symbols) while the anion curves (red dashed lines) follow the cation simulation results (black symbols). Our extended MSA $\mu_+$ and $\mu_-$ split much more correctly, although there is sometimes a minimum for $\mu_-$ that is not present in the simulation results. This maybe where Eq.\ (\ref{mui}) breaks down, or it maybe that Eq.\ (\ref{mui}) is too early to reveal a minimum that occurs at higher concentrations in the simulations. Currently, the simulated concentrations are not high enough to make the distinction, so this will need to be explored in future work.

\begin{figure}[htp]
    \centering
    \includegraphics[width=3in]{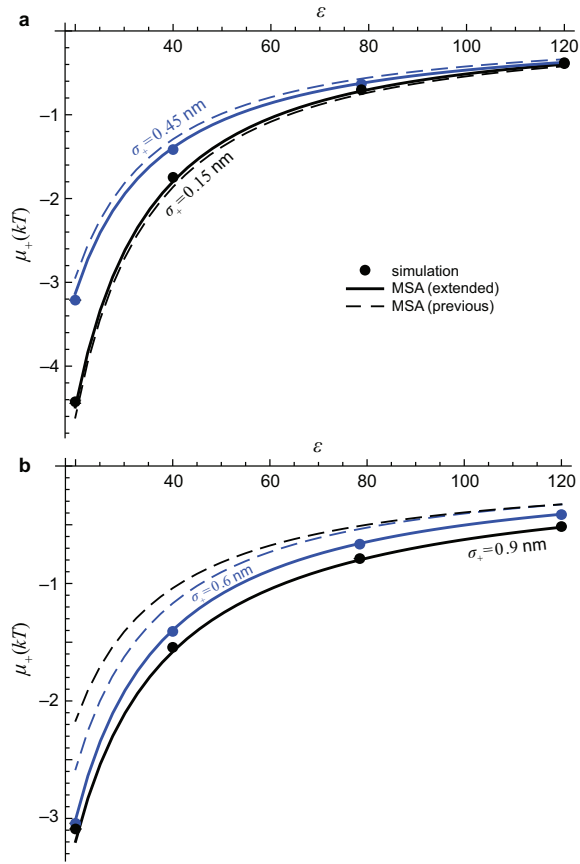}
    \caption{Comparison of our extended MSA $\mu_+$ from Eq.\ (\ref{mui}) (solid lines) to both the standard MSA formulation Eq.\ (\ref{muiMSA}) (dashed lines) and MC simulations (symbols) as dielectric constant $\varepsilon$ varies. In all panels, only cation results are shown. The anion diameter is always $0.3$ nm, while the cation diameter $\sigma_+$ is indicated on each set of curves. Each panel shows the results for two $1:1$ electrolytes with $\rho_+=1$ M, one indicated with black lines/symbols and the other with blue lines/symbols.
    }
    \label{fig:compare2}
\end{figure}

Fig.\ \ref{fig:compare2} shows that our Eq.\ (\ref{mui}) also works extremely well at different dielectric constants $\varepsilon$. Here, $\mu_+$ versus $\varepsilon$ is shown for both small cations (Fig.\ \ref{fig:compare2}a) and large cations (Fig.\ \ref{fig:compare2}b). As in Fig.\ \ref{fig:compare1}, the previous MSA formulations and our extended version give very similar results when the cations are small (Fig.\ \ref{fig:compare2}a). However, the two are quite different when the cations are large compared to the anions (Fig.\ \ref{fig:compare2}b). In fact, the previous MSA formulation's error becomes larger with larger size asymmetry. This can be seen in Fig.\ \ref{fig:compare2}b by noting that the dashed black line is much farther from the solid black line (the $0.9$ nm cation case) than the dashed and solid blue curves are from each other (the $0.6$ nm cation case).

\section{Conclusions}

We have extended the standard MSA theory to fully account for individual ion species excess chemical potentials (Eq.\ (\ref{mui})). To verify the accuracy of this theory, in the figures we directly compared against simulation for at least $28$ unique electrolytes (i.e., different combinations of $z_+:z_-$, $\sigma_+$, and $\varepsilon$) over a very wide range of concentrations ($1$ $\mu$M to $>1$ M). These included challenging combinations of trivalents ($z_+=3)$, very large ions ($\sigma_+=0.9$ nm), and small and large dielectric constants ($20\le \varepsilon \le 120$). The excellent results of our extended MSA give us confidence not only that our two derivations (main text and Appendix \ref{Appneuback}) are correct, but also that Eq.\ (\ref{mui}) is useful for real-world applications whenever the primitive model of electrolytes is a valid model (e.g., ionic liquids).

\begin{acknowledgments}
Research reported in this publication was supported by the National Heart, Lung, and Blood Institute of the National Institutes of Health under award number R01HL057832 to DG. The content is solely the responsibility of the authors and does not necessarily represent the official views of the National Institutes of Health.
\end{acknowledgments}

\section*{Author declarations}
\subsection*{Conflict of interest}
The authors have no conflicts to disclose.

\section*{Data availability}
The data that support the findings of this study are available from the corresponding author
upon reasonable request.

\appendix

\section{MSA of electrolytes in a uniform neutralizing background}
\label{Appneuback}

On a macroscopic scale, an electrolyte of uniform ion densities has to fulfill charge neutrality. If this is obtained by adding a uniform background charge, then the chemical potential of single ions may be derived in a slightly different way than in Sec.\ \ref{chempotls}. For such a background, a high density of charged point particles spread uniformly throughout the electrolyte can be used. Let these particles have density $\rho_0$, valence $z_0$, and hard core diameter $\sigma_0=0$. Then, the limit $\rho_0\rightarrow \infty$ and $z_0\rightarrow0$ is considered with $z_0 \rho_0$ finite. The background will interact with the regular ions, but they will remain uncorrelated. With high density they will give a corresponding high contribution to pressure, but this does not influence correlations in the limit considered.

The usual MSA solution for charged hard spheres cannot be applied directly to this system. The reason is that it is restricted to additive hard spheres. To that point, charges also have to stay outside the hard cores from other charged particles. But, as we will argue and find below, the MSA system may be transformed to represent a system with a uniform background.

The internal energy per particle $u_i$ of component $i$ from the MSA solution of charged hard spheres (Eqs.\ (\ref{ui})--(\ref{ustar})) is
\begin{equation}
  \beta {{u}_{i}}=\frac{{{\alpha }^{2}}}{4\pi }{{z}_{i}}{{N}_{i}}+{{z}_{i}}\beta {{u}^{*}}
\end{equation}
with $N_i$ given by Eq.\ (\ref{Ni}) and
\begin{equation}
\label{ustarapp}
    	\beta {{u}^{*}}=-\frac{{{\alpha }^{2}}}{24 }\sum\limits_{l}{{{\rho }_{l}}\sigma _{l}^{2}\left( {{N}_{l}}{{\sigma }_{l}}+\frac{3}{2}{{z}_{l}} \right)}.
\end{equation}
The internal energy density of the ionic fluid is \cite{hoye21}
\begin{eqnarray}
\beta E&=&\sum_i\beta \rho_i u_i = \frac{\alpha^2}{4\pi}\sum_i\rho_i z_i N_i\\
&=&-\frac{\alpha^2}{4\pi}\left(\Gamma\sum_i\frac{\rho_i z_i^2}{1+\Gamma\sigma_i}+\frac{\pi}{2\Delta}\Omega P_n^2\right).
\label{103}
\end{eqnarray}
where $\Delta$ is given by Eq.\ (\ref{Delta}), $P_n$ by Eq.\ (\ref{Pn}), and $\Omega$ by Eq.\ (\ref{Om}).
Charge neutrality requires
\begin{equation}
\sum_i z_i \rho_i=0,
\label{104}
\end{equation}
by which the quantity $u^*$ does not contribute to $E$. The Helmholtz free energy per unit volume $A$ is given by \cite{hoye21}
\begin{equation}
\beta A=\beta E+\frac{\Gamma^3}{3\pi}.
\label{105}
\end{equation}
Since
\begin{equation}
\frac{\partial(\beta A)}{\partial \Gamma}=0,
\label{106}
\end{equation}
Eq.\ (\ref{105}) gives the thermodynamic relation $\partial (\beta A)/\partial \beta=E$, as required.  (We verify Eq.\ (\ref{106}) below.)

With relation (\ref{106}) the chemical potentials can be found as
\begin{equation}
\beta\mu_i=\frac{\partial(\beta A)}{\partial \rho_i}=\frac{\partial(\beta E)}{\partial \rho_i}=\beta u_{0i}+\delta\mu_{0i},
\label{108}
\end{equation}
where
\begin{equation}
\beta u_{0i}=\frac{\alpha^2}{4\pi} z_i N_i 
\end{equation}
and
\begin{equation}
\delta\mu_{0i}=-\frac{\alpha^2}{4\pi}\eta\sigma_i\frac{\eta \sigma_i^2(\Gamma\sigma_i-2)+3z_i}{3(1+\Gamma\sigma_i)}.
\label{109}
\end{equation}
Due to charge neutrality, the $u^*$ term does not appear here and has no influence. 

Now consider the system with a uniform background. The background is then also present inside the hard sphere ions and effectively adds a charge to them. The influence of this charge on its surroundings is the same as for a point charge at the center of each ion. This is a well-known property of Coulomb interaction. 
However, there will an additional interaction between the given point charge with valence $z_l$ and the part of the background inside the hard core diameter $\sigma_l$. This clearly is proportional to the last term of the quantity $u^*$ in Eq.\ (\ref{ustarapp}). Moreover, the first term of $u^*$ represents the volume of a sphere of diameter $\sigma_l$. The background charge inside it is proportional to this volume. Thus, there is reason to expect that the quantity $u^*$ has something to do with the neutralizing background.

Next, we show that the system with a background charge can be transformed into an effective MSA problem. This is done by adding the background charge inside each sphere to its given point charge with valence $z_i$. Then, the particle gets an effective valence of
\begin{equation}
z_{ei}=z_i+\frac{\pi}{6}z_0\rho_0\sigma_i^3
\label{110}
\end{equation}
for ion species $i=1,2,3,\hdots ,n$ where $n$ is the number of components apart from the background. Since part of the background has been added to the particle charges, its average density from the remaining ions has been reduced in magnitude (with fixed $z_0$, the $\rho_0$ can be greater or less than $0$). Its effective density becomes
\begin{equation}
\rho_{e0}=\rho_0\left(1-\frac{\pi}{6}\sum_l\sigma_l^3\rho_l\right).
\label{111}
\end{equation}
Charge neutrality requires
\begin{equation}
\sum_i \rho_i z_i+\rho_0z_0=0.
\label{112}
\end{equation}
Eqs.~(\ref{110})--(\ref{112}) then give
\begin{equation}
\sum_i \rho_i z_{ei}+\rho_{e0}z_0=0.
\label{113}
\end{equation}
Thus, we have obtained an effective MSA system where charge neutrality is fulfilled.  The main part of its internal energy is given by Eq.~(\ref{103}) with $z_i \mapsto z_{ei}$. It can be noted that there is no contribution from the background when $z_0\rightarrow0$. The situation is the same for the quantities $\Omega$ and $P_n$ since $\sigma_0=0$.

As indicated earlier, there is an additional contribution to the internal energy. This is the interaction between the ionic point charges $z_i$ and the background inside each hard sphere ion. With charge density that follows from Eq.~(\ref{110}), this Coulomb energy $\Delta u_i$ is
\begin{eqnarray}
\beta \Delta u_i&=&\frac{\alpha^2}{4\pi} 4\pi \int\limits_0^{\sigma_i/2}\frac{1}{r}z_i\rho_0z_0 r^2\,dr\\
&=&\frac{\alpha^2}{4\pi}\frac{\pi}{2} z_i \sigma_i^2 z_0 \rho_0.
\label{114}
\end{eqnarray}
This gives the additional internal energy density of
\begin{eqnarray}
\beta \Delta E=\sum_i\beta\rho_i\Delta u_i=\frac{\alpha^2}{4\pi}\frac{\pi}{2}\sum_l(z_l\rho_l\sigma_l^2 )z_0\rho_0.
\label{115}
\end{eqnarray}
One sees that this term is similar to the $z_l$ term of Eq.~(\ref{ustarapp}). Altogether the internal energy is
\begin{equation}
E=E_e+\Delta E
\label{116}
\end{equation}
where $E_e$ is Eq.\ (\ref{103}) with $z_i \mapsto z_{ei}$. With this modification the Helmholtz free energy density (\ref{105}) will remain the same. The reason is that $z_{ei}$ and $\Delta E$ do not contain the parameter $\Gamma$, and so the partial derivative (\ref{106}) still vanishes. Thus, the chemical potentials are given by
\begin{equation}
\beta\mu_i=\frac{\partial(\beta A)}{\partial\rho_i}=\frac{\partial(\beta E)}{\partial\rho_i}.
\label{117}
\end{equation}
By differentiation where $z_{ei}$ and $\Delta E$ are kept constant, result (\ref{108}) is recovered. Then, one can differentiate with respect to $z_{ei}$ to get
\begin{eqnarray}
\frac{\partial(\beta E)}{\partial z_{el}}
&=&- 2\frac{\alpha^2}{4\pi}\left[\Gamma\frac{\rho_l z_{el}}{1+\Gamma\sigma_l}+\frac{\pi}{2\Delta}P_n\frac{\rho_l\sigma_l}{1+\Gamma\sigma_l}\right]\\
&=&2\frac{\alpha^2}{4\pi}\rho_l N_l.
\label{118}
\end{eqnarray}
With charge neutrality (\ref{112}), expression (\ref{110}) can be rewritten as
\begin{equation}
z_{el}=z_l-\frac{\pi}{6}\sigma_l^3\sum_k z_k \rho_k.
\label{119}
\end{equation}
Then,
\begin{eqnarray}
\beta\delta\mu_{1i}&\equiv&\sum_l\frac{\partial(\beta E)}{\partial z_{el}}\frac{\partial z_{el}}{\partial \rho_i}\\
&=&-2\frac{\alpha^2}{4\pi}z_i \frac{\pi}{6}\left(\sum_l\rho_l\sigma_l^2 N_l\sigma_l\right).
\label{120}
\end{eqnarray}
Likewise, with charge neutrality, we have
\begin{eqnarray}
\beta\delta\mu_{2i}&\equiv&\frac{\partial (\beta\,\Delta E)}{\partial \rho_i}\\
&=&-2\frac{\alpha^2}{4\pi}z_i\frac{\pi}{6}\left(\sum_l\rho_l\sigma_l^2\frac{3}{2}z_l\right),
\label{121}
\end{eqnarray}
where the limit of zero background charge is used (i.e., $z_0 \rho_0 \rightarrow 0$). Adding these together gives
\begin{equation}
\beta\delta\mu_{1i}+\beta\delta\mu_{2i}=-2z_i u^*
\label{122}
\end{equation}
with $u^*$ given  by Eq.\ (\ref{ustarapp}). Thus, with expressions (\ref{108}), (\ref{109}), and (\ref{122}) the individual species chemical potentials become
\begin{equation}
\beta\mu_i=\beta u_{0i}+\beta\delta\mu_{0i}-2z_i u^*.
\label{123}
\end{equation}
This is the identical to our main result, Eq.\ (\ref{mui}).

Lastly, we verify Eq.\ (\ref{106}).
By partial differentiation one first finds with expression (\ref{103})
\begin{eqnarray}
\label{130}
\frac{\partial(\beta E)}{\partial \Gamma}&=&-\frac{\alpha^2}{4\pi} \sum_i\frac{\rho_i z_i^2}{(1+\Gamma\sigma_i)^2}\\
&&-\frac{\alpha^2}{4\pi}\frac{\pi}{2\Delta}\left(2P_n\frac{\partial}{\partial\Gamma}(\Omega P_n)-\frac{\partial\Omega}{\partial\Gamma}P_n^2\right).
\nonumber
\end{eqnarray}
With some algebra, one further finds
\begin{equation}
\frac{\partial(\Omega P_n)}{\partial \Gamma}=- \sum_i\frac{\rho_i\sigma_i^2 z_i}{(1+\Gamma\sigma_i)^2},
\end{equation}
\begin{equation}
\frac{\partial\Omega}{\partial \Gamma}=-\frac{\pi}{2\Delta}\sum_i\frac{\rho_i \sigma_i^4}{(1+\Gamma\sigma_i)^2},
\label{131}
\end{equation}
\begin{eqnarray}
\frac{\partial(\beta E)}{\partial \Gamma}&=&-\frac{\alpha^2}{4\pi} \sum_i\frac{\rho_i \left(z_i^2-2\eta\sigma_i^2 z_i+\eta^2\sigma_i^4\right)}{(1+\Gamma\sigma_i)^2} \\
&=&-\frac{\Gamma^2}{\pi},
\label{132}
\end{eqnarray}
where the last equality follows from expression (\ref{Gamma}) for $\Gamma$. Thus, with expression (\ref{105}) for $A$ we have
\begin{equation}
\frac{\partial(\beta A)}{\partial \Gamma}=\frac{\partial(\beta E)}{\partial \Gamma}+\frac{\Gamma^2}{\pi}=0,
\label{133}
\end{equation}
which is Eq.~(\ref{106}).

\section{Derivation of $u^*$}
\label{Appustar}
Since H\o ye and Blum \cite{hoye78} only stated the result for the internal energy per particle (Eq.\ (\ref{ui})) with $N_i$ defined by Eq.\ (\ref{Ni}) and $u^*$ defined by Eq.\ (\ref{ustar}), we offer a short derivation here. We use Blum's review of the MSA derivation in Ref.~\onlinecite{blum80} as our guide.

We start with Eq.\ (\ref{uidef}), the definition of excess internal energy beyond the mean-field contribution. With Coulombic interactions,
\begin{eqnarray}
\int h_{ij}(r) \beta \psi_{ij}(r,0) d \mathbf{r}
& = & z_i z_j \alpha^2 \int t h_{ij}(t) dt\\
& = & z_i z_j \frac{\alpha^2}{2\pi} J_{ij}(0)
\end{eqnarray}
for $J_{ij}(r)$ defined in Eq.\ (\ref{Jij}).
Therefore,
\begin{equation}
    \beta u_i=\frac{\alpha^2}{4\pi} z_i B_i
\end{equation}
where
\begin{equation}
    B_i = \sum_k z_k \rho_k J_{ki}(0).
\end{equation}

By Eq.\ (B.40) of Ref.~\onlinecite{blum80} (and the equivalent Eq.\ (2.31) of Ref.~\onlinecite{blum75}), the $B_i$ are given implicitly by
\begin{equation}
\label{NiBi}
    N_i = B_i + \frac{\pi}{4\Delta}
    \left[ \chi_2 + \frac{2}{3} \sum_k \rho_k \sigma_k^3 B_k
    \right]
\end{equation}
where $\Delta$ is from Eq.\ (\ref{Delta}) and
\begin{equation}
    \chi_2 = \sum_k z_k \rho_k \sigma_k^2.
\end{equation}
Since we know $N_i$ from Eq.\ (\ref{Ni}), we can solve for the $B_i$. The matrix associated with this linear system has a unique structure that allows for an analytic solution. Specifically, Eq.\ (\ref{NiBi}) in matrix form is
\begin{equation}
    \left( I + \begin{pmatrix} \xi_1 & \cdots & \xi_n \\ \vdots & & \vdots \\ \xi_1 & \cdots & \xi_n \end{pmatrix} \right) \Vec{B}
    = \Vec{N} - \begin{pmatrix} \chi \\ \vdots \\ \chi \end{pmatrix}
\end{equation}
where $I$ is the identity matrix, $\chi = \pi \chi_2 / 4 \Delta$ and $\xi_k = \pi \rho_k \sigma_k^3 / 6 \Delta$. The solution of this is
\begin{equation}
    B_i = \frac{r_i + r_i \sum_k \xi_k - \sum_k r_k \xi_k}{1+\sum_k \xi_k}
\end{equation}
for $r_k = N_k - \chi$. Substituting in for $r_k$, $\chi$, and $\xi_k$ gives Eq.\ (\ref{ui}) after some algebra.

\section{Evaluation of the sum $\sum_k\rho_k a_k q'_{ik}$}
\label{secA2}

The sum $\beta u_{0i}=(z_i/4\pi)\sum_k\rho_k a_k q'_{ik}$ in Eq.\ (\ref{sum}), via the definition of $q_{ik}'$ in Eq.\ (\ref{qp}), is given by
\begin{equation}
    \beta u_{0i} = \frac{z_i}{4\pi}\left\{\frac{\pi}{\Delta}\left[\sigma_i S_1+\left(1+\frac{\pi}{2\Delta}\zeta_2\sigma_i\right)S_2\right]-\frac{2\Gamma^2}{\alpha^2}a_i S_3\right\}
\end{equation}
for the $S_i$ defined below.

To evaluate the sum
\begin{equation}
\label{ui0}
    S_1=\sum_k\rho_k a_k,
\end{equation}
we rewrite the $a_k$ given by Eq.~(\ref{ai}) as 
\begin{equation}
a_k=\frac{\alpha^2}{2\Gamma}\left[z_k-\frac{z_k\Gamma\sigma_k}{1+\Gamma\sigma_k}-\frac{\pi}{2\Delta}P_n\sigma_k^2+\frac{\pi}{2\Delta}\frac{\Gamma\sigma_k^3}{1+\Gamma\sigma_k}\right].
\label{64}
\end{equation}
Then,
\begin{eqnarray}
S_1 & = & \frac{\alpha^2}{2\Gamma}\left[0-\Gamma\Omega P_n-\frac{\pi}{2\Delta} P_n\zeta_2+P_n(\Omega-1)\Gamma\right]\\
& = & \frac{\alpha^2}{2\Gamma}\left[-\frac{\pi}{2\Delta} P_n\zeta_2-P_n\Gamma\right],
\label{65}
\end{eqnarray}
with $P_n$ and $\Omega$ given by Eqs.~(\ref{Pn}) and (\ref{Om}).

With $a_k$ of form (\ref{ai}), 
\begin{eqnarray}
S_2&=&\sum_k\rho_k \sigma_k a_k\\
&=&\frac{\alpha^2}{2\Gamma}[\Omega P_n-P_n(\Omega-1)]\\
&=&\frac{\alpha^2}{2\Gamma}P_n.
\label{66}
\end{eqnarray}

Finally, by Eq.\ (14) of Ref.~\onlinecite{hoye21},
\begin{equation}
S_3=\sum_k\rho_k a_k^2=\alpha^2.
\label{67}
\end{equation}

Adding everything together, we find
\begin{eqnarray}
\nonumber
\beta u_{0i}&=&\frac{z_i}{4\pi}\left\{\frac{\pi}{\Delta}\left[\sigma_i S_1+\left(1+\frac{\pi}{2\Delta}\zeta_2\sigma_i\right)S_2\right]-\frac{2\Gamma^2}{\alpha^2}a_i S_3\right\}\\
\nonumber
&=&\frac{z_i}{4\pi}\left[\frac{\pi\alpha^2}{2\Delta\Gamma}(-\Gamma\sigma_i+1)P_n-2\Gamma^2 a_i\right]\\
\nonumber
&=&\frac{z_i}{4\pi}\left[\frac{\pi\alpha^2}{2\Delta\Gamma}(-\Gamma\sigma_i+1)-2\Gamma^2\alpha^2\frac{z_i-\eta\sigma_i^2}{2\Gamma(1+\Gamma\sigma_i)}\right]\\
&=&\frac{\alpha^2}{4\pi}z_i\left(N_i+\frac{\eta}{\Gamma}\right).
\label{68}
\end{eqnarray}
Here, the last term of this result is the same as the last sum of Eq.~(\ref{sum}), while the first one is $\beta u_i-z_i\beta u^*$. 

For the analysis of the remaining terms of (\ref{sum}), we only give a brief outline. We note that the last terms of Eqs.~(\ref{ctildemethod2}) and (\ref{cmethod2}) differ only by $\beta u_{0i}$ plus another term; this follows from Eqs.~(35) and (43) of Ref.~\onlinecite{hoye21}. This other term, by that reference's Eq.~(A17), is a ${{z}_{i}}\cdot \text{const}\text{.}$ term that was neglected in Ref.~\onlinecite{hoye21}. 
The remaining terms, except for another ${{z}_{i}}\cdot \text{const}\text{.}$ term in its Eq.\ (A20), of its Eq.~(44) (which computes $\beta \mu_i - \beta u_i$) were included in their $\mu_i^{\textrm{MSA}}$. This latter ${{z}_{i}}\cdot \text{const}\text{.}$ term, like $z_i \delta^*$ of Eq.~(\ref{ziterm}), exactly compensates the last sum of Eq.~(\ref{sum}), from which Eq.\ (\ref{mui}) follows.


\end{document}